\documentclass[prl,aps,twocolumn,superscriptaddress,longbibliography,noeprint]{revtex4-2}
  
\usepackage {graphicx}
\usepackage {amssymb}
\usepackage {amsmath}
\usepackage{color}

\newcommand{\reffig}[1]{Fig.~\ref{#1}}
\newcommand{\refeq}[1]{Eq.~(\ref{#1})}
\newcommand{\refeqs}[2]{Eqs.~(\ref{#1})-(\ref{#2})}

\newcommand{\vect}[1]{\mathrm{\mathbf{#1}}} 
\newcommand{\real}{\textrm{Re}\,}
\newcommand{\imag}{\textrm{Im}\,}

\newcommand{\bra}[1]{\ensuremath{\left\langle#1\right|}}
\newcommand{\ket}[1]{\ensuremath{\left|#1\right\rangle}}
\newcommand{\bracket}[2]{\ensuremath{\left\langle#1\right.\left|\, #2\right\rangle}}

\newcommand{\INVremark}[1]{}
 \newcommand{\INVtodo}[1]{}

\begin{document}

\title{All-optical attoclock for imaging tunnelling wavepackets}

\author{I. Babushkin}
\affiliation{Institute for Quantum Optics, Leibniz
  Universit\"at Hannover, Welfengarten 1, 30167 Hannover, Germany}
\affiliation{Cluster of Excellence PhoenixD (Photonics, Optics, and Engineering – Innovation Across Disciplines), Welfengarten 1, 30167 Hannover, Germany}
\affiliation{Max Born Institute, Max Born Str. 2a, 12489
  Berlin, Germany}
\author{A. J. Galan}
\affiliation{Max Born Institute, Max Born Str. 2a, 12489
  Berlin, Germany}
\author{J. R. C. Andrade}
\affiliation{Max Born Institute, Max Born Str. 2a, 12489
  Berlin, Germany}
\author{A. Husakou}
\author{F. Morales}
\author{M.~Kretschmar}
\author{T. Nagy}
\affiliation{Max Born Institute, Max Born Str. 2a, 12489
  Berlin, Germany}
 \author{V. Vai\v{c}aitis} 
  \affiliation{Laser Research Center, Vilnius University,
    Saul\.{e}tekio 10, Vilnius LT-10223, Lithuania}
  \author{L. Shi}
  \affiliation{Institute for Quantum Optics, Leibniz
  Universit\"at Hannover, Welfengarten 1, 30167 Hannover, Germany}
  \author{D. Zuber}
  \affiliation{Institute for Quantum Optics, Leibniz
  Universit\"at Hannover, Welfengarten 1, 30167 Hannover, Germany}
\affiliation{Cluster of Excellence PhoenixD (Photonics, Optics, and Engineering – Innovation Across Disciplines), Welfengarten 1, 30167 Hannover, Germany}
\author{L. Berg\'e}
\affiliation{CEA, DAM, DIF, F-91297 Arpajon, France}
\affiliation{Université Paris Saclay, CEA, LMCE, 91680 Bruyères-le-Châtel, France}
\author{S. Skupin}
\affiliation{Institut Lumi\`ere Mati\`ere, UMR 5306 Universit\'e Lyon 1 -- CNRS, Universit\'e de Lyon, 69622 Villeurbanne, France}
\author{I.A.Nikolaeva}
\affiliation{Faculty of Physics,
  Lomonosov Moscow State University, 1 Leninskie gory, Moscow 119991, Russia}
\affiliation{Lebedev Physical Institute of Russian Academy of Sciences, 53 Leninskiy prospect, Moscow 119991, Russia}
\author{N.A.Panov}
\affiliation{Faculty of Physics,
  Lomonosov Moscow State University, 1 Leninskie gory, Moscow 119991, Russia}
\affiliation{Lebedev Physical Institute of Russian Academy of Sciences, 53 Leninskiy prospect, Moscow 119991, Russia}
\author{D.E.Shipilo}
\affiliation{Faculty of Physics,
  Lomonosov Moscow State University, 1 Leninskie gory, Moscow 119991, Russia}
\affiliation{Lebedev Physical Institute of Russian Academy of Sciences, 53 Leninskiy prospect, Moscow 119991, Russia}
\author{O.~G. Kosareva}
\affiliation{Faculty of Physics,
  Lomonosov Moscow State University, 1 Leninskie gory, Moscow 119991, Russia}
\affiliation{Lebedev Physical Institute of Russian Academy of Sciences, 53 Leninskiy prospect, Moscow 119991, Russia}
  \author{A. N. Pfeiffer}
\affiliation{Institute of Optics and Quantum Electronics, Abbe Center of Photonics, Friedrich Schiller University, Max-Wien-Platz 1, 07743
Jena, Germany}
  \author{A. Demircan}
\affiliation{Institute for Quantum Optics, Leibniz
  Universit\"at Hannover, Welfengarten 1, 30167 Hannover, Germany}
\affiliation{Cluster of Excellence PhoenixD (Photonics, Optics, and Engineering – Innovation Across Disciplines), Welfengarten 1, 30167 Hannover, Germany}
\affiliation{Hannover Centre for Optical Technologies, Nienburger Str. 17, 30167 Hannover, Germany}
  \author{M. J. J. Vrakking}
\affiliation{Max Born Institute, Max Born Str. 2a, 12489
  Berlin, Germany}
\author{U.~Morgner}
\affiliation{Institute for Quantum Optics, Leibniz
  Universit\"at Hannover, Welfengarten 1, 30167 Hannover, Germany}
\affiliation{Cluster of Excellence PhoenixD (Photonics, Optics, and Engineering – Innovation Across Disciplines), Welfengarten 1, 30167 Hannover, Germany}
\affiliation{Hannover Centre for Optical Technologies, Nienburger Str. 17, 30167 Hannover, Germany}
\author{M. Ivanov}
\affiliation{Max Born Institute, Max Born Str. 2a, 12489
  Berlin, Germany}

\maketitle

\textbf{Recent experiments on measuring time-delays during tunnelling
  of cold atoms through an optically created potential barrier are
  reinvigorating the controversial debate regarding possible
  time-delays during light-induced tunnelling of an electron from an
  atom. Compelling theoretical and experimental arguments have been
  put forward to advocate opposite views, confirming or refuting the
  existence of finite tunnelling time delays. Yet, such a delay,
  whether present or not, is but a single quantity characterizing the
  tunnelling wavepacket; the underlying dynamics are richer. Here we
  propose to augment photo-electron detection in laser-induced
  tunnelling with detection of light emitted by the tunnelling
  electron -- the so-called Brunel radiation. Using a combination of
  single-color and two-color driving fields, we identify the
  all-optical signatures of the re-shaping of the tunnelling
  wavepacket as it emerges from the tunnelling barrier and moves away
  from the core. This reshaping includes not only an effective
  time-delay but also time-reversal asymmetry of the ionization
  process, which we describe theoretically and observe experimentally.
  We show how both delay and reshaping are mapped on the polarization
  properties of the Brunel radiation, with different harmonics
  behaving as different hands of a clock moving at different speeds.
  The all-optical detection paves the way to time-resolving optical
  tunnelling in condensed matter systems, e.g. tunnelling across
  bandgaps in solids, on the attosecond time-scale. }

\section{Introduction}

Tunnelling of an electron through the potential barrier created by an
oscillating electric field and the binding potential of the core is a
key resource in attosecond science \cite{corkum07} and is at the heart
of high harmonic generation and high harmonic spectroscopy
\cite{baker06,smirnova09,bruner15, torres10,pedatzur15}. Generally,
high harmonic generation is associated with radiative recombination
following the return of the laser-driven electron to the parent ion.
Yet, even when the electron does not return to the core, harmonic
radiation is still emitted. Often referred to as the ``Brunel
radiation'' or ''Brunel harmonics'' \cite{brunel90,
  babushkin17,balciunas13}, it is associated with bursts of current
triggered by laser-induced tunnelling and is ubiquitous in atoms,
molecules, and solids \cite{lanin17,silva19,jurgens20,juergens21}.
Sub-cycle tunnelling bursts occur every time the oscillating electric
field goes through a maximum. Both odd and even harmonics emerge when
the incident laser field contains both fundamental radiation and its
second harmonic, including the so-called '0-th' order Brunel harmonic
corresponding to terahertz (THz) emission, see e.g.
\cite{kim08b,babushkin11,babushkin17}. To produce THz radiation, the
waveshape must be at least asymmetric in respect to the change
$\vect E\to-\vect E$.
The origin of this radiation can be traced already in a classical
picture: According to Maxwell's equations, a change of the electric
current $d\vect J/dt$ produces radiation. A small volume emits, in a
small time interval $dt$, Brunel radiation
$\delta \vect E_\mathrm{Br}\propto d \vect J$ produced by the change
of free electron current $d \vect J \propto \rho(t)\vect E(t)dt$ due
to acceleration of electrons with density $\rho(t)$ in the driving
field $\vect E(t)$ in the situation when the $\rho(t)$ is time-dependent
\cite{geissler99,juergens21}. Additional terms may arise due to, for
instance, nonzero birth velocity of electrons (see Methods), but the
above expression is rather good approximation for the lowest harmonics
\cite{babushkin10,zhang20}.

Here we show that different Brunel harmonics generated in elliptically
polarized single-color and two-color laser fields provide a detailed
picture of light-induced tunnelling. Our approach to imaging the
ionization dynamics is distinctly different from the currently used
attoclock approaches based on detecting photo-electrons
\cite{eckle08a,eckle08,huismans11,shafir12,kienberger04,pfeiffer11,
  landsman14,yakaboylu14,torlina15,ni16,camus17,sainadh19,eicke20}. It
allows us to introduce a complementary, all-optical measurement
protocol, which should enable one to extend measurements of tunnelling
dynamics into the bulk of solids, where photo-electron spectroscopy is
not readily available.

\begin{figure*}
\includegraphics[width=\textwidth]{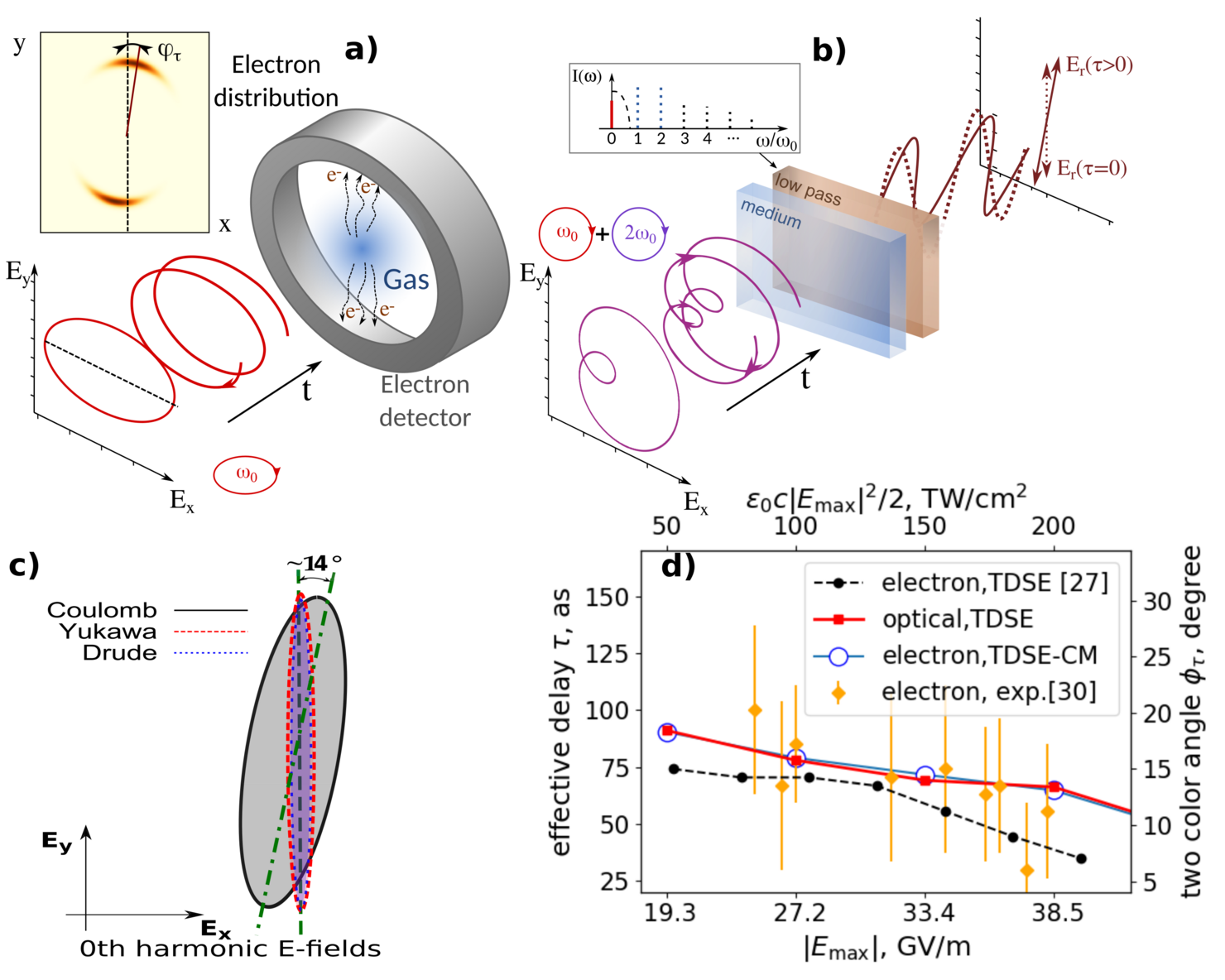}
\caption{ \label{fig:base} Photoelectron (a) and optical (b-d)
  attoclock. In (a), the angle-resolved photo-electron spectrum
  generated by the driving field (red line) is registered at a
  detector, revealing attosecond delays and deflections of electronic
  wavepacket shifting its maximum by $\phi_\tau$, interpreted as an
  effective delay $\tau=\phi_\tau/\omega_0$. In (b), this delay is
  probed, instead of electrons, by the 0th-order Brunel radiation in a
  two-color field (magenta line). Such harmonic generated in gas or
  solid (blue box) can be selected by a low-pass filter (brown box).
  It is nearly linearly polarized (brown line) and rotated, having the
  same effective $\tau$ as in (a). The inset shows schematically the
  harmonic response spectrum, including the pump (blue) and harmonic
  of interest (red). (c) Polarization state of the 0th harmonic for
  the equivalent intensity $\epsilon_0 c |E_\mathrm{max}|^2/2=150$
  TW/cm$^2$, obtained from TDSE simulations with Yukawa potential,
  Coulomb potential, and using the simple-man classical Drude model. (d)
  Effective delay $\tau$ as a function of intensity obtained from the
  polarization rotation of the 0th harmonic as determined by TDSE
  simulations and compared to the results of the photo-electron
  attoclock: the center of mass position of the electronic wavepacket
  (TDSE-CM), as well as the simulations of \cite{torlina15} (where the
  plane $z=0$ instead of center of masses was analyzed) and the experiment
  of \cite{sainadh19}. }
\end{figure*}

In the strong field regime, the conventional photo-electron attoclock
\cite{eckle08a,eckle08,ivanov14} measures the deflection in the
photo-electron spectrum generated by a few-cycle, nearly circularly
polarized pulse, with respect to the pulse polarization ellipse.
Interpretation of these observations uses mapping between the instant
at which the electron ``escapes the tunnelling barrier'' and its final
velocity at the detector (see \reffig{fig:base}a). This
interpretation, and the possible time-delays associated with the
ionization, are a subject of active discussions
\cite{yakaboylu14,landsman14,landsman15,ni16,teeny16,camus17,kaushal15,torlina15,eicke19},
just as the general problem of tunnelling times, see e.g. the latest
beautiful experiment on this subject \cite{ramos20}. The difficulty
stems from the need to deconvolve ionization and post-ionization
dynamics in the presence of the electron--core interaction, which
deflects the outgoing electron and thus distorts the mapping between
the tunnelling event and the measurement at a faraway detector.
Moreover, possible under-the-barrier delays and the post-barrier
electron velocity make the one-to-one mapping from the state at
detector and emission time impossible 
\cite{teeny16,camus17}. Optical techniques, relying on measuring
recombination-based high harmonic emission \cite{pedatzur15, bruner15}
instead of electron detection, face an additional challenge: the need
to decouple the ionization and photo-recombination steps.

Therefore, it is advantageous to place the "observation point" as
close to the tunnel exit as possible. Brunel radiation is an ideal
candidate: it is produced by the current generated by the tunneling
electron right near the barrier exit. We find that already the lowest,
0th-order Brunel harmonic, encodes the attosecond-scale delays and the
wavepacket deflection by the core potential. The information is
contained in the rotation of the polarization ellipse of this
harmonic. We prove this analytically and numerically by comparing the
proposed new protocol against the standard attoclock setup.

Moreover, we find theoretically and confirm experimentally that
higher-order Brunel harmonics encode the temporal profile of the
ionization burst, providing several connected ``clock hands'' rotating
at different speed. Using the third-order Brunel harmonic generated in
helium, we are able to reconstruct the dynamical reshaping of the
electronic wavepacket during ionization. Interestingly, we find
asymmetry in the ionization dynamics relative to the instantaneous
peak of the driving electric field. We attribute this effect to pre-
and post-ionization electron dynamics.

\section{Physical principle and theoretical analysis}

We first validate the key idea behind the all-optical attoclock: the
vectorial properties of the emitted light, such as the rotation of the
polarization ellipse, are determined by the vectorial properties of
the current generated by the tunnelling electron and therefore reflect
the tunnelling dynamics.

We consider two field arrangements. In the first one, intense
circularly polarized infrared (IR) pump with frequency $\omega_0$ is
combined with its co-rotating second harmonic $2\omega_0$, see
\reffig{fig:base}b. This generates a total electric field with one
well-pronounced maximum per cycle. The orientation of this maximum,
here set along the $x$-axis, is controlled by the relative phase
between the two colors and provides the reference direction for the
optical attoclock. In the second arrangement, such reference direction
is provided by the major axis of the single-color elliptically
polarized driving field.

\subsection{All-optical measurement of strong-field ionization 
delays}

We begin with the first arrangement. The nonlinear response contains
even and odd harmonics $n\omega_0$, including the "zeroth-order"
harmonic, where the signal is dominated by the Brunel radiation
\cite{meng16,tulsky18,tailliez20}. Let a classical free electron be
injected by strong-field ionization into the atomic continuum with
some (not necessarily zero) velocity and be accelerated in the laser
field and the potential of the core. The electric field associated
with the Brunel radiation, emitted by this electron, reads
$\vect E_{\mathrm{Br}}(\omega)\propto \int \frac{d{\vect p}(t)}{dt}
e^{i\omega t}dt$, with $\vect p$ being the electron momentum. For the
0-th harmonic $\omega\to 0$, and
$\vect E_{\mathrm{Br}}(0)\propto \vect p(t\to\infty)$. Thus, the
polarization direction of the 0-th Brunel harmonic tracks the final
velocity of the electron after the interaction, i.e. performs the same
measurement as the photo-electron attoclock, but by all-optical means.
Note that this conclusion is independent of assumptions regarding the
ionization process. While exactly zero frequency cannot be observed in
the experiment, measurements in the low-frequency (THz) range are
feasible and discussed below.

Our conclusion also holds within the fully quantum-mechanical
analysis, see Methods. In this case, $\vect E_{\mathrm{Br}}(0)$ tracks
the center of mass of the electronic wavepacket. Specifically,
\begin{equation}
\phi[\vect E_{\mathrm{Br}}(0)]=\phi_\tau =\int \phi P(\phi) d\phi, \label{eq:EBR0}   
\end{equation}
where $\phi[\vect E_{\mathrm{Br}}(0)]$ denotes the azimuthal angle
formed by the vector $\vect E_{\mathrm{Br}}(0)$ in the $xy$-plane,
orthogonal to the light propagation direction $z$, measured clockwise
from the positive $y$ direction;
$P(\phi)=\int |\psi(t\to\infty,p,\phi,\theta)|^2p^2\sin\theta
dpd\theta$ is the photo-electron distribution integrated over the
momentum and the polar angle, and $\psi(t,p,\phi,\theta)$ is the
electron wavefunction in $\vect p$-representation. In the
photo-electron attoclock, the angle $\phi_\tau$ is mapped onto the
effective ionization time-delay $\tau$ as $\phi_\tau=\omega_0 \tau$,
where $\omega_0$ is the carrier defining the "standard attoclock
mapping" $\mathcal{M}$,
$\mathcal M(\phi_\tau)=\phi_\tau/\omega_0=\tau$. For our two-color
field configuration, \reffig{fig:base}b, the mapping is slightly
different (see Methods):
$\tau =\mathcal M(\phi_\tau) \simeq \phi_\tau/\omega'_0$, where
$\omega'_0 = 4/3\omega_0$. This mapping should not be confused with
the physical interpretation of $\tau$, which requires the analysis of
the deflection of the tunnelling electron in the core potential,
possible energy gain during optical tunnelling and possibly non-zero
initial electron velocity when exiting the barrier
\cite{yakaboylu14,landsman14,landsman15,ni16,teeny16,camus17,kaushal15,torlina15,eicke19}.

Putting together \refeq{eq:EBR0} and the definition of $\mathcal M$
for the two-color field configuration, the polarization vector of the
0-th Brunel harmonic is predicted to be
\begin{equation}
    \vect E_{\mathrm{Br}}(0) \sim (\sin(\omega'_0 \tau),\cos(\omega'_0\tau)), \label{eq:EBRtau}
\end{equation}
where $\tau$ is the same effective ionization delay 
as obtained from the photo-electron spectrum.  

We now verify this conclusion by using ab-initio time-dependent
Schr\"odinger equation (TDSE) simulations to compute the radiated
field. We start with the short-range Yukawa potential, which allows us
to remove the effects of the long-range Coulomb tail of the core. In
this case, the TDSE simulations show that the THz radiation is indeed
nearly linearly polarized and that the major axis of its polarization
ellipse is orthogonal to the major axis of the Lissajous figure of the
driving two-color field, \reffig{fig:base}b,c.
Thus, we confirm that, for a short-range potential supporting a single
bound state, both the photo-electron attoclock and the optical
attoclock measure zero ionization delays. This result also coincides 
 with the result of classical simple-man Drude model, where
\refeq{eq:drude-time} was used with ionization rate $\rho(t)$
calculated using Yudin-Ivanov formula \cite{yudin01}. The finite
ellipticity seen in the TDSE results is due to the finite spectral
width of the response.

For the Hydrogen atom, TDSE simulations show rotation of the
polarization ellipse of the 0-th harmonic, \reffig{fig:base}c. Figure
\ref{fig:base}d compares the time-delay extracted from this rotation
with the numerical results of \cite{torlina15} and experimental
results of \cite{sainadh19} for the photo-electron attoclock, as well
as the positions of the centers of mass of the electronic wavepackets.
The data points in Fig. 1d show the average polarization as discussed
in Methods. 
This average is weighted with the spectral energy density,
and coincides with the rotation angles at $\omega\to 0$ (see
Supplementary). Again, we confirm that the two-color all-optical
attoclock demonstrates essentially the same effective time delay
$\tau$ as the single-color photo-electron one, especially if we
compare the rotation angles with the center of mass of the electronic
wavepacket, as our theory suggests. 
The broader polarization ellipse for the Coulomb potential is due to
the broader spectrum of the THz response and the fact that higher
frequencies are more elliptically polarized (see Supplementary). 

\subsection{Extending optical attoclock using 
higher-order Brunel harmonics}

\begin{figure*}
  \includegraphics[width=17cm]{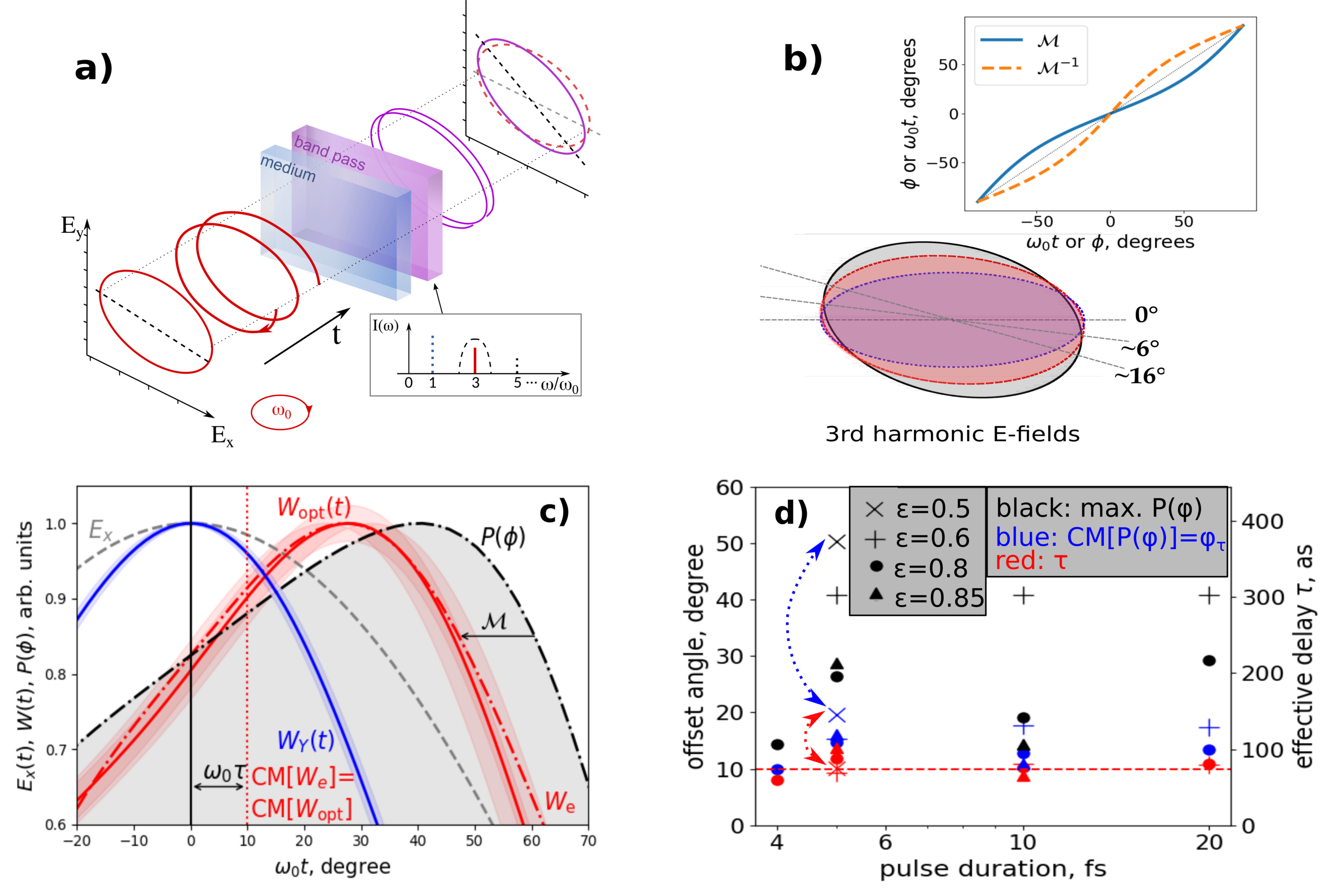}
  \caption{ \label{fig:sy} Extracting photoionization information from
    higher order Brunel harmonics for the Hydrogen atom. In (a), a
    single-color elliptically polarized pump (red line) produces the
    3rd Brunel harmonic (magenta line), with the polarization state
    (ellipticity and polarization direction) encoding the sub-cycle
    dynamics of the ionization process. Inset shows schematically the
    harmonic response spectrum, including the pump (blue) and harmonic
    of interest (red). (b) Polarization of the 3rd harmonic for the
    single-color driving pulse ($|E_\mathrm{max}|=36.3$ GV/m,
    $\epsilon=0.6$, $I=280$ TW/cm$^2$) calculated using TDSE with
    Coulomb and Yukawa potentials as well as simple-man Drude model
    (denotations as in Fig. 1c). Inset shows the attoclock mapping
    $\mathcal{M}$ and its inverse. (c) Optical reconstruction of the
    ionization dynamics ($W_\mathrm{opt}(t)$, red solid line) compared
    to the reconstruction from the photoelectron spectrum ($W_{e}(t)$,
    red dot-dashed line) for the Coulomb potential. The estimation of
    error is given in Methods. Optical reconstruction for the Yukawa
    potential ($W_Y(t)$, blue line) is also presented. Black
    dot-dashed line shows the electron wavepacket distribution
    $P(\phi)$. The attoclock delay $\omega_0\tau$ given by the the
    position of the center of mass (CM) of $W_{e}(t)\approx W_{\mathrm{opt}}(t)$ is also
    indicated (red vertical dotted line). (d) The positions of the
    maxima of the electronic spectra $P(\phi)$ (black markers), their
    CM(blue markers) and the effective delays $\tau=\mathcal{M}(\phi)$
    reconstructed from the photo-electron spectra (red markers) for
    different FWHM pulse durations and ellipticities $\epsilon$ of the
    driving field. Horisontal red line shows the attoclock delay
    extracted optically using the two-color configuration (Fig. 1) for
    the corresponding peak intensity. Asymmetry of the wavepacket
    reveals itself from the different positions of the maximum and CM
    of the photo-electron spectra (see the dotted arrows tracing these
    for an exemplary case of $\epsilon=0.5$). }
\end{figure*}

The attoclock mapping $\phi_\tau\to\tau$ establishes the relation
between only two numbers, $\phi_\tau$ and $\tau$. We now extend the
optical attoclock beyond the measurement of a single parameter to
access the sub-cycle ionization yield. To this end, we consider the
second field arrangement with a single-color, elliptically polarized
field, see \reffig{fig:sy}a, and focus on the rotation of the
polarization ellipse of the higher-order Brunel harmonics.

Consider again the simple model of a classical electron extracted with
zero velocity. The Brunel radiation, given by the electrons born with
ionization rate $W(t)$ and producing free current $\vect J(t)$ is
described by a simple-man (so-called Drude) model
\cite{kim09,babushkin10}:
\begin{equation}
   \label{eq:drude-time}
   \vect E_\mathrm{Br}(t) \propto d\vect J/dt \propto \vect E(t)\rho(t) \propto \vect E(t) \int_{-\infty}^t W(t')dt',
 \end{equation}
where $\rho(t)$ is the free electron density.  

On the other hand, we can also use the photo-electron spectrum and the
attoclock mapping between the polar detection angle $\phi$ of
electrons, measured at infinity, to define an effective ionization
time $t$ and ionization rate $W_e(t)$,
\begin{equation}
   \label{eq:Mp}
   W_e(t)  = P(\phi) d\phi/dt, \, \phi = \mathcal M^{-1}(t).
\end{equation}
Note that such definitions have nothing to do with the interpretation
of the ionization delays in either atto-clock measurement.

As shown in the Methods section using the R-matrix approach
\cite{torlina12,torlina15}, the simple-man equation \refeq{eq:Mp}
remains fully valid also beyond the classical approximation, at least
for few lowest order harmonics, if we assume $W(t)=W_e(t)$. Thus, as
long as the measured harmonics are associated with the Brunel
radiation, observing $\vect E_\mathrm{Br}(t)$ gives all-optical access
to $P(\phi)$, imaging the tunneling wavepacket by all-optical means.
Below we confirm this conclusion by comparing $W(t)$ obtained
optically and directly from the photo-electron distribution.

To reconstruct $W(t)$ from the Brunel radiation, we rewrite
\refeq{eq:drude-time} in the frequency domain:
\begin{equation}
  \label{eq:drude-model-freq}
   \vect E_{\mathrm{Br},n}  \propto \frac{i}{\omega_0}\sum_j
 \frac{\vect E_j  W_{n-j}}{n-j},
\end{equation}
where $W_k$ and $\vect E_j$ are the Fourier components of the
ionization rate and of the driving field, respectively, and
$\vect E_{\mathrm{Br},n}$ are $n$-th Brunel harmonics.

Already the polarization state of the 3rd harmonic gives rich
information about the ionization dynamics. For the two co-rotating
circular fields used above, the polarization of the 3rd harmonic is
very close to circular, and thus no information can be extracted from
its polarization direction since the latter is undefined. For the
single-color elliptical pump (with ellipticity $\epsilon\ne 1$), see
\reffig{fig:sy}a, Eq (\ref{eq:drude-model-freq}) yields
\begin{equation}
  \label{eq:E3}
  \vect E_{{\rm Br},3} \propto \left(1+r/2,
    i\epsilon(1-r/2)\right), \, 
  r=W_4/W_2. 
\end{equation}

\refeq{eq:E3} shows that the polarization state of the 3rd harmonic
depends critically on the 2nd and 4th harmonics of the ionization
burst $W(t)$. Most importantly, it allows us to determine whether the
ionization burst is centered at the field maximum and symmetric around
this center, or delayed and/or asymmetric. In the first case, the
ratio $r=W_4/W_2$ is real-valued and the rotation of the polarization
of the third harmonic is zero. In the second case, the ratio $r$ is
complex-valued and the polarization is rotated. Thus, extracting the
phase of $r$ from the rotation of $\vect E_{{\rm Br},3}$ and comparing
it with the ionization delay allows one to identify the presence of
asymmetry in the ionization burst $W(t)$ with respect to its maximum.
Additional information is, of course, provided by the higher-order
harmonics and the phase of $r'=W_6/W_4$, $r''=W_8/W_6$ etc. In the
experiment, we focus on $\vect E_{{\rm Br},3}$.

To evaluate this approach, we first make TDSE simulations for the
Hydrogen atom in our single-color field configuration (Fig. 2a) and
observe the polarization of the 3rd harmonic, see Fig. 2b. As
expected, for the Yukawa potential the rotation of
$\vect E_{{\rm Br},3}$ is noticeably smaller than for the Hydrogen
atom and is different from the one expected from the time-delay
measured by the 0-th harmonic of the two-color optical attoclock.
Together, the information extracted from both measurements is already
sufficient to reconstruct $W(t)$ defined by \refeq{eq:Mp} up to 4th
harmonic,
\begin{equation}
  \label{eq:wi}
  W_{\rm opt} (t) = W_0 + W_2 \cos(2\omega_0t + \delta_2) + W_4 \cos(4\omega_0t +
  \delta_4),
\end{equation}
where, we changed the notations as $2|W_j|\to W_j$,
$\arg(W_j)\to \delta_j$ with respect to \refeq{eq:drude-model-freq}.
The subscript in $W_{\rm opt} (t)$ stresses the optical route for the
reconstruction. The coefficients $W_2$, $W_4$, $\delta_2$, $\delta_4$
are reconstructed uniquely if we assume that the center of mass of
$W_{\rm opt} (t)$ appears at $t=\tau$ (see details in Methods).

The resulting $W_{\rm opt} (t)$ is compared in \reffig{fig:sy}c with
$W_{\rm e} (t)$ extracted directly from the photo-electron spectra
$P(\phi)$ (red solid and red dot-dashed curves, respectively). We find
excellent agreement between both, proving the viability of the
all-optical method.

Figure \ref{fig:sy}c shows a clear asymmetry in $W(t)$ and $P(\phi)$
with respect to time reversal, i.e., the field maximum, in the case of
the Coulomb potential. This asymmetry disappears completely for the
Yukawa potential, proving that the Coulomb tail not only introduces
the time-delay, but also distorts the outgoing wavepacket. While in
retrospect such conclusion appears transparent, hindsight is always
20/20: to the best of our knowledge, this is the first time when such
asymmetry has been both reported and quantified. The presence of both
non-trivial ionization phase and the distortion of the outgoing
electronic wavepacket is important not only for the Brunel, but also
for the recollision-based harmonics in molecules, where it will
manifest as channel-dependent ionization phase associated with
different ionic states between ionization and recombination, and thus
different effective core potentials. Such channel-dependent ionization
phase has indeed been confirmed experimentally in \cite{bruner16}.
Interestingly, our simulations (see \reffig{fig:sy}d) also show that
the asymmetry becomes small for nearly circularly polarized and
few-cycle pulses, suggesting that asymmetry could be also related to
intermediate excitations building up inside the potential well,
expected in the regime of intermediate Keldysh parameters
\cite{ivanov05}. This interpretation is further confirmed by the
pronounced Freeman resonances in the rotation angle of
$\vect E_{\mathrm{Br},3}$ as a function of laser intensity; Besides,
for high input intensities, saturation effects start to play a role  (see
Supplementary information for the case of He atom).

We note that the ionization delay $\tau$ extracted from the
photo-electron spectra is nearly independent of the ellipticity and
pulse duration and coincides with that extracted optically, using the
two-color configuration, see \reffig{fig:sy}d. We see thus that the
ionization delay is, as we can judge, waveform-independent, in
contrast to the deflection of the electronic cloud.

\section{Imaging ionization dynamics of Helium}

   \begin{figure*}
\includegraphics[width=\textwidth]{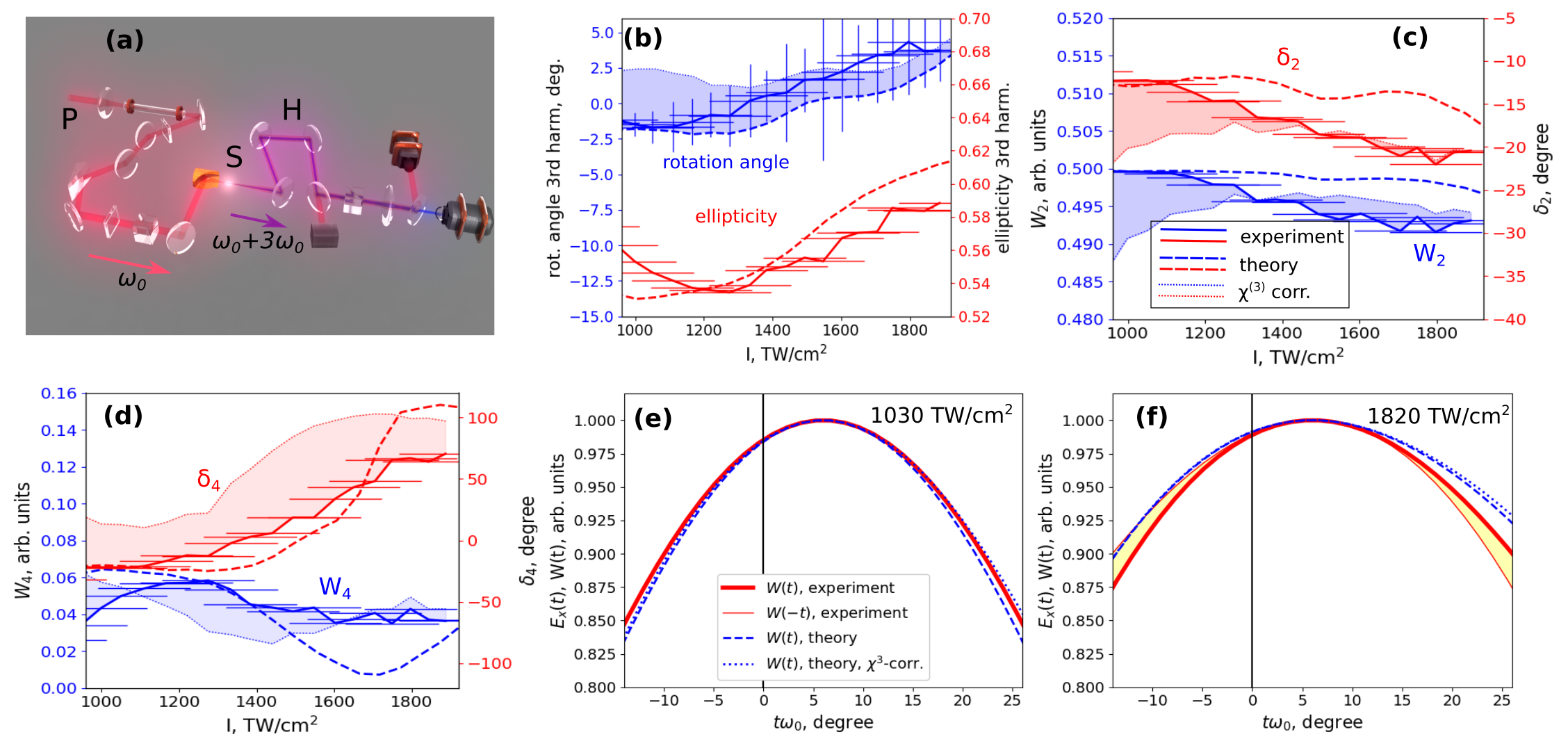}
\caption{ \label{fig:exp} Experimental reconstruction of the subcycle
  ionization dynamics in He, compared to the theoretical simulations.
  (a) Experimental setup (see Methods for more details): A 800-nm,
  43-fs long elliptically-polarized pump pulse (P) is focused into a
  plasma spot (S) where 3rd harmonic (H) is generated. The
  polarization components are carefully separated and detected. (b)
  Experimentally measured rotation of the long axis of the 3rd
  harmonic with respect to the long axis of the pump (solid blue
  curve, left axis) and ellipticity of the 3rd harmonic (solid red
  curve), in comparison with the results of TDSE calculations (dashed
  blue and red lines). The shaded region shows uncertainty introduced
  by influence of bound-bound transitions. Further errors are
  discussed in Methods. (c),(d) Reconstruction of the coefficients
  $W_i$, $\delta_i$ describing ionization dynamics as given by
  \refeq{eq:wi} (with denotations as in (b)). (e),(f) Ionization dynamics
  $W(t)$ reconstructed for an exemplary intensities $I=1030$~TW/cm$^2$~(e) 
  and $I=1820$~TW/cm$^2$~(f) according to experiment (thick red
  solid lie) and theory (blue dashed line). Blue dotted line indicates
  the theoretical curve with bound-bound-transitions correction
  included. Thin red solid line indicates the time inverted $t\to -t$
  version of $W(t)$ and yellow-dashed region between red lines thus
  indicates the asymmetry of $W(t)$ (in (e) the asymmetry is almost
  invisible). }
\end{figure*}

We now move to the experimental results. The predicted rotation of the
polarization ellipse of the nonlinear response is confirmed by the
experimental measurements with setup depicted in \reffig{fig:exp}a.
Here we report the rotation angle and the ellipticity of the third
harmonic generated by a strong pulse with ellipticity $\epsilon=0.61$
focused in helium. The experimental approach is described in detail in
the Methods section. The polarization of the pump was measured and
found rotate by a negligible angle of less than 0.5 degrees upon the
propagation, thus excluding the influence of plasma-induced
birefringence. We note nevertheless, that the Brunel mechanism of 3rd
harmonic generation competes with the $\chi^{(3)}$-based one. Since
those two mechanisms give contributions with relative phase $\pi/2$,
this results in additional polarization rotation (see Supplementary
Material). The correction induced by the $\chi^{(3)}$ nonlinearity is
shown as a blue shaded region above the theoretical curves in
\reffig{fig:exp}. Its impact decreases with intensity and becomes
almost negligible above $I\approx 1500$ TW/cm$^2$.

The experimentally measured intensity-dependent parameters of the
polarization ellipse (rotation angle and ellipticity) are shown in
\reffig{fig:exp}b, compared with the results of TDSE simulations for
the helium atom. The details regarding the error bars are described in
the Methods. Experiment and simulation are in a very good agreement.
The polarization of the 3rd harmonic is rotated by nearly 5 degrees as
the intensity increases from 1000 to 2000~TW/cm$^2$, while the
ellipticity moderately increases from around 0.56 to 0.62. Combining
this information with the theoretical method developed above, we are
able to reconstruct the subcycle ionization dynamics, taking the
strong-field ionization delay $\tau$ of He from the experiments
performed in \cite{eckle08}.

The coefficients $W_2$, $W_4$, $\delta_2$, $\delta_4$, as defined in
\refeq{eq:wi} and reconstructed from our measurement, are shown in
\reffig{fig:exp}(c,d) as a function of the pump intensity. We find
that $W_2$ and $W_4$ remain nearly constant as intensity increases,
whereas the angles $\delta_2$ and $\delta_4$ experience significant
variation. The fact that $W_i$ barely change shows that ionization
remains ``equally fast'' for the range of intensities in the
experiment. On the other hand, the growth of $|\delta_i|$ indicates
the changing asymmetry of the effective sub-cycle ionization rate
$W(t)$ with respect to its maximum. In \reffig{fig:exp}(e,f), $W(t)$
is visualized for two values of the pump intensity, showing that the
asymmetry increases with increasing intensity. While the asymmetry
remains fairly small even for large intensities, its presence in the
measurements is clearly seen from \reffig{fig:exp}(c,d): the measured
values of $|\delta_i|$ deviate substantially from zero, well beyond
the experimental error bars. Thus, even small asymmetries lead to
large deviations of the rotation angle, which is well detected by the
all-optical method.

\section{Conclusions and Outlook}

In conclusion, we have established a firm quantitative link between
measuring the photo-electron spectra in strong-field ionization and
measuring the Brunel radiation generated by electrons on their way to
the continuum, revealing the reshaping of the electronic wavepacket
during laser-induced tunnelling. Such reshaping is mapped on the
effective ionization delays and the asymmetry of the instantaneous
ionization rate with respect to the peak of the field, as it is imaged
by the Brunel harmonics.
According to this, Brunel harmonics in THz and UV
  contain signatures of attosecond- and sub-angstrom- scale electron
  dynamics, deeply sub-wavelength for these frequencies. Remarkably, we
observe that the ionization delay is largely independent on the
driving field waveshape, in contrast to the spatial reshaping of the
electronic wavepacket.
We attribute the origin of the ionization asymmetry to the dynamics of
the electron wavepacket during and after tunnelling and, for high
intensities, saturation effects. Our experiment in helium shows that
the all-optical method is very sensitive even to small asymmetries of
such kind. This demonstrates promising capabilities of the proposed
new protocol for imaging tunnelling, which allows one to explore
attosecond-scale wavepacket reshaping in systems where photo-electron
detection is not readily available. One of such systems is the bulk
solids, where the detection of light is much simpler than the
detection of electrons and where attosecond-scale delays were recently
reported \cite{hofmann15,sommer16,hassan16}. We foresee that Brunel
harmonics of yet higher order will allow resolving electron dynamics
even closer to the core.
  
\section{Methods}

\begin{footnotesize}

\subsection{Experimental setup}

The laser used in the experiments is a Ti:Sapphire-based regenerative
amplifier (Spectra Physics Spitfire) delivering pulses with 3 mJ of
energy and the spectrum supports a 43 fs of pulse duration at the
driving wavelength of 800 nm. The beam was delivered to the
experimental setup via an evacuated hollow-core fibre with 70\%
transmission. This ensured a clean beam profile with an
$\text{M}^{2}<1.1$ and a 5.5 mm beam diameter. The delivered energy
was selected using a pair of thin-film polarizers in combination with
a broadband half-wave plate. To further attenuate the pulses and
ensure the polarization purity of the electric field a fused silica
wedge was used in Brewster's angle and the s-polarized pulse was used
in the experiment. This allowed a tuning of the pulse energy between
40 and 225 $\mu$J with a high linear polarization contrast. An
achromatic quarter-wave plate covering the whole range of the laser
spectrum was used to convert the linear polarization into elliptical.
The polarization was analyzed using a Glan-Laser calcite polarizer.
The measured ellipse had an intensity ratio of 0.372 between the minor
and major axes. The light was then guided into a vacuum chamber
containing a parabolic gold mirror with a focal distance of 100 mm.
After the focus the beam was recollimated using an Aluminium curved
mirror and guided out of the chamber using further Aluminium coated
mirrors. A dichroic beam-splitter was used to separate the
third-harmonic from most of the fundamental. The IR light rejected by
the dichroic beam splitter was directed onto a power meter in order to
analyze power fluctuations during the measurements. The reflected
light was used in the polarization characterization portion of the
experiment. The polarization state of the third harmonic was analyzed
using an $\alpha$-BBO Glan-Laser polarizer (ThorLabs GLB10-UV) mounted
on a home-built electronic rotation stage. Detection was performed
after a dielectric HR mirror for 266 nm light allowing further
separation of the UV component from the IR. The 800~nm transmitted
through the 266 nm HR mirror was detected on a Silicon photodiode
while the UV was measured using a Silicon Carbide (SiC) photodiode.
The SiC detector is not sensitive to visible or infrared radiation,
which allowed a measurement where the background is not influenced by
any co-propagating IR light. All three measurement devices were
connected to analogical boxcar integrators and the signals were then
digitized using a 16-bit ADC connected to an arduino nano. This setup
allowed to characterize the polarization state of the UV elliptical
light, and to compare the direction of its long axis to that of the
exciting IR field. The chamber was evacuated and subsequently filled
with He at a pressure of 1.3 bar. No other nonlinearities from the
optical components were observed.

\subsection{Pump fields in time and frequency representation, and
the attoclock mapping $\mathcal M$}
    
We consider driving pulses propagating in $z$-direction, with electric
field in OXY plane: $\vect E = \{E_x,E_y\}$ and vector potential
$\vect A \{A_x,A_y\}$, $\vect A(t)=-\int_{-\infty}^t \vect E(t')dt'$
in the form:
\begin{gather}
\label{eq:field1}
  A_{x}(t) = \frac{E_0}{\omega_0}e^{-2 \ln 2 \,t^2/\mathcal{T}^2} \cos{\omega_0 t},\\
  A_{y}(t) = -\epsilon \frac{E_0}{\omega_0}e^{-2 \ln 2 \,t^2/\mathcal{T}^2} \sin{\omega_0 t},
  \label{eq:field1a}
\end{gather}
where, in our simulations $\omega_0$ corresponds to the carrier
wavelength of 0.8 $\mu$m, $E_0$ is the field
amplitude. 
The pulse duration $\mathcal{T}$ and the ellipticity $\epsilon$
were chosen to be 
20 fs and $0.61$
for helium (Fig. 3). For Fig. 2c, $\mathcal{T}$ and  $\epsilon$ were 10 fs and $0.6$, correspondingly.
For the two-color field we have
\begin{gather}
\label{eq:field20}
  A_{x}(t) = -\frac{E_0}{\omega_0}e^{-2 \ln 2 \,t^2/\mathcal{T}^2} \left(\sin{\omega_0 t} + \frac12\sin{2\omega_0 t} \right),\\
  A_{y}(t) = -\frac{E_0}{\omega_0}e^{-2 \ln 2 \,t^2/\mathcal{T}^2} \left(\cos{\omega_0 t}+\frac12\cos{2\omega_0 t}\right),
  \label{eq:field2}
\end{gather}
with the same $\omega_0$,
and $\mathcal{T}=20$ fs.

The attoclock mapping $\mathcal M^{-1}$ which maps time $t$ to the
direction of the vector potential $\vect A(t)$, that is,
$\phi=\mathcal{M}^{-1}(t)\equiv\arctan{(A_y(t)/A_x(t))}$, with $\phi$
defined such that it respects the angle definition after Eq. (1), is
then determined as follows: For the pulse given by
\refeqs{eq:field1}{eq:field1a} we have (neglecting slow envelope
effects):
\begin{equation}
  \label{eq:M1}
  t = \mathcal M(\phi) \equiv \frac{1}{\omega_0}\arctan{\left(\epsilon \tan(\phi)\right)}.
\end{equation}
For small $\phi$ can be approximated as
$ \mathcal{M}(\phi)\approx\phi/\omega'_0$,
$\omega'_0=\omega_0/\epsilon$. For nearly circularly polarized pulses
($\epsilon \approx 1$) this expression approaches the ``traditional''
mapping $\tau=\mathcal{M}(\phi)=\phi/\omega_0$. Both $\mathcal M$ and
$\mathcal M^{-1}$ are shown in the inset to Fig. 2b.
For \refeqs{eq:field20}{eq:field2}, the mapping $\mathcal{M}(\phi)$ is difficult to obtain analytically, but its inverse  $\mathcal{M}^{-1}(t)$ is given by:
\begin{equation}
  \label{eq:M2}
  \phi = \mathcal M^{-1}(t) \equiv \arctan{\left(\frac{2\cos{(\omega_0t)}+\cos{(2\omega_0t)}}{2\sin{(\omega_0t)}+\sin{(2\omega_0t)}}\right)},
\end{equation}
which is rather good approximated by 
$\mathcal M^{-1}(\tau) \approx 4\omega_0\tau/3$. That is,
$\tau = \mathcal M(\phi) \approx \phi/\omega'_0$,
$\omega'_0=4\omega_0/3$.

\subsection{Jones vector formalism}

In deriving the equations for the Brunel harmonics response, we used
the Jones vector representation for the field harmonics
$\vect E(t) = \sum_n \vect E_ne^{-in\omega_0t}$. The waveshape
described in \refeqs{eq:field1}{eq:field2}, can be easily translated
into $\vect E_n$. For instance, a single-color ﬁeld with carrier
(fundamental) frequency $\omega_0$ deﬁned by
$E_x = E_0\cos{(\omega_0 t)}$, $E_y= -E_0\sin{(\omega_0 t)}$ (which
corresponds to the fundamental frequency component of the 2-color
field) we can easily obtain using the Jones formalism as:
$\vect E_1 = E_0/2\cdot(1,-1/i)$,
$\vect E_{-\vect 1} = E_0/2\cdot(1,1/i)$ where the first and second
component of the vector are its $x$- and $y$- components,
respectively. In the main text of the manuscript, the insignificant
pre-factor $E_0/2$ is removed since only the polarization state
matters.

\subsection{TDSE simulations} 

The TDSE equation has the well-known form (we use atomic units here  and
in the following subsection):
\begin{equation}
    i \partial_t \psi(\vect r,t) = H\psi(\vect r,t), \, H=(\vect
    p+\vect A(t))^2/2+V(r), 
    \label{eq:1a}
\end{equation}
where $\psi(\vect r, t)$ is the wavefunction of the electron depending
on space $\vect r$ (with $r\equiv |\vect r|$) and time $t$, $H$ is the
electron's Hamiltonian, with $\vect p$ being the momentum operator,
$\vect A(t)$ the vector potential of the external (laser) field with
the electric field strength $\vect E(t) = -\partial_t \vect A$. The
simulation box size used for the integration in the two color case
(Fig.1) was 500 au. In this case, to ensure the numerical convergence,
simulations were repeated with both uniform and logarithmic spatial
grids. For the single color case, the photo-electron spectrum (PES) in
Fig. 2 was extracted by means of the iSURF method
\cite{morales16isurf} using projection to Volkov states. The box size
was taken to be up to 300 au in this case. The convergence of PES was
also tested by projecting to the Coulomb states, obtained in a
simulation with a box of 3000 au.

\subsection{Reconstruction of harmonic response from the TDSE simulations}

The field $\vect E_r$ resulting from the atomic response was
calculated as
$\vect E_r = g\partial_{tt}\vect p =g\partial_t \vect j$, where
$\vect p= \bra{\psi}\vect r\ket{\psi}$ is the atomic polarization,
$\ket{\psi}$ is the electron wavefunction, $\vect j=\partial_t\vect p$
is the current assigned to $\vect p$, and $g$ is a constant depending
on the position of the observer, which we assume equal to unity for
simplicity. Note that this response contains not only Brunel but also
all other nonlinearity mechanisms. For the results shown in
\reffig{fig:exp}, the relevant quantities (polarization direction,
ellipticity) were averaged in the range
$(0.75n\omega_0,1.25n\omega_0)$ for the $n$th harmonic except for
$n=0$. For $n=0$ (Fig. 1), the average was made in the range
$(0,0.25\omega_0)$, that is from zero to approx. 100 THz. In all
cases, the average was performed using weighting with the intensity
distribution. Examples of the spectral dependencies for 0th harmonic
are shown in the Supplementary (Fig. 8). 

\subsection{The  Drude-like model from R-matrix theory} 

In this section we derive the Drude-like model
\refeq{eq:drude-model-freq} using the framework of the analytic
R-matrix approach \cite{torlina12, torlina15}. In this approach, the
electron wavefunction is written as:
$\psi(\vect r,t) = \sum_{t_s,\vect p} \bracket{\vect r}{\vect p}
R_{\vect p} e^{iS(\vect p,t, t_s)}$ where $R_{\vect p}$ encodes the
angular structure of the initial state, $t_s$ is a (complex-valued)
stationary time (see below), and $S = S_{f} + S_C$,
$S_{f}= 1/2\int_{t_s-i\kappa^{-2}}^t(\vect p+\vect A(t'))^2dt' -
I_p(t_s-t)$,
$S_C=\int_{t_s-i\kappa^{-2}}^tV(\vect r (\vect p, \tau,t_s))d\tau$,
where $\kappa=\sqrt{2I_p}$ and $I_p$ is the ionization potential of
the atom. The ionization time $t_s$ is inferred via the stationary
phase approach from the equation $\partial_{t_s}S=0$. $t_s$ is
different from the ionization time $t_0$ in a short range potential
defined from $\partial_{t_0}S_f=0$, and hereupon we express
$\tau=\real{(t_s-t_0)}$. According to \cite{torlina15}, the value of
$\tau$ can be obtained as a perturbation from the relation
$-\tau \partial_{t_s t_s}S_f=\partial_{t_s}S_C$. The approach of
\cite{torlina15} can be also seen in the following way: by
neglecting second and higher order derivatives of $S_C$ as well as
third and higher order derivatives of $S_f$, in the vicinity of $t_0$
we obtain
$S(\vect p,t,t_s)\approx S_f(\vect p,t,t_0+\tau)+C(t,\vect p)$. That
is, by changing variables $t+\tau\to t$ we transform (approximately)
$S\to S_f +\mathrm{const}(t_0)$. Here, we extend this approach, by noting
that there always exists a (in general complex) function
$\tau(t,\vect p,t_0)$,
that makes the mapping mentioned above exact:
\begin{equation}
S(\vect p,t,t_s) = S_f(\vect p,t,t_0+\tau(t,\vect p,t_0)) +C(t,\vect
p)\label{eq:s-sf}
\end{equation}
for
arbitrary $t_0$. That is, introducing an effective delay $\tau$ should
effectively map
the Coulomb-driven dynamics after the exit of the barrier to
the dynamics without the Coulomb tail. Formally speaking, this mapping is
general enough to connect arbitrary functions  $S$ and $S_f$ which are
analytic everywhere except at isolated points. 
This leads us to the following expression for the free wavefunction
\begin{equation}
\psi(\vect r,t) = \sum_{t_0,\vect p}\bracket{\vect r}{\vect
  p}R'_{\vect p} e^{iS_f(\vect p,t, t_0+\tau(t,\vect p,
  t_0))},\label{eq:phi-gen}
\end{equation}
with $R'_\vect p = R_\vect p e^{iC(t,\vect p)}$.
The physical meaning of $\tau$ requires that $\tau(t,\vect p,t_0)\to0$
as $t\to t_0$, since if we consider electrons directly after the
tunnel exit, they have not seen the Coulomb tail yet. On the other
hand, as the modulus of $T = t-t_0$ becomes large,
$\tau(t,\vect p,t-T)$ should become independent of $T$ (meaning that
at large times after the electrons were born they do not see the
Coulomb tail anymore). That is, $\tau(t,\vect p,t-T)$ approaches some
limiting function $\tau(t,\vect p)$ at large $|T|$. In this situation,
the expression for free electron density $\rho(t)$ can be most
straightforwardly and transparently derived assuming a constant
$ C(t,\vect p)\approx = C$ and only time dependent
$\tau(t,\vect p)\approx \tau(t)$ at least for those $\vect p$ which
contribute mostly to \refeq{eq:phi-gen} (see below the discussion how
these assumptions can be relaxed). In this case we obtain:
\begin{equation}
\rho(t) = C_0 \rho_f(t-\tau(t)),\label{eq:del-rho}
\end{equation}
where $C_0=e^{-2\imag C}$, and $\rho_f(t)$ is the free electron
density in a short range potential (that is, assuming $C=0$, $\tau=0$
in \refeq{eq:phi-gen}).
Neglecting $\partial_t \tau$, we also obtain that the ionization rate
$W(t) \propto d\rho/dt$ which obeys the same rule:
$W(t) = C_0W_f(t-\tau(t))$, and $W_f(t)$ is the corresponding
ionization rate for the short-range potential. Note that,
independently from the simple derivation above, \refeq{eq:del-rho}
provides a general transformation to map the two electron densities
$\rho(t)$ and $\rho_f(t)$, except for pathological cases, if
$\rho(t\to-\infty)=\rho_f(t\to-\infty)=0$ holds. In this case,
$\tau(t)$ can be calculated directly from $\rho(t)$ and $\rho_f(t)$.
That means, \refeq{eq:del-rho} is valid also beyond the
simplifications used in the derivation above. In that more general
case, the expression relating the delay $\tau$ in \refeq{eq:del-rho}
and in \refeq{eq:s-sf} will be more involved.

To complete the derivation of our Drude-like model, we take into
account that since the mapping \refeq{eq:s-sf} has been applied, we
have to consider electrons propagating free from the action of the
Coulomb tail. Neglecting quantum interference effects in the
electronic wavepacket and the term responsible for the photon
absorption  \cite{geissler99,jurgens20,juergens21}, we can write the
current $\vect J(t)$ formed by free electrons at large enough times
$t$ as a sum of ``elementary currents'' from electrons born at
particular times:
$\vect J(t) = \int_{-\infty}^t\delta \vect J(t,t')dt'$, with
$\delta \vect J(t,t')dt'=-n_0C_0W_f(t'-\tau(t'))(\vect A(t)-\vect
A(t')+\vect v_0(t')))dt'$ represent the current from electrons born
between $t'$ and $t'+dt'$ with initial velocity $\vect v_0$ and atomic
density $n_0$. Differentiating this expression and taking into account
\refeq{eq:del-rho}, we obtain:
\begin{equation}
  \label{eq:brunel-v0}
  \frac{d\vect J}{dt} = n_0\left(\rho(t)\vect E(t) - W(t)\vect v_0(t)\right).
\end{equation}
Using the relation $d \rho \approx W(t)dt$, this expression can be reformulated as:
\begin{equation}
  \label{eq:brunel-v0a}
  d\vect J \propto \rho(t)\vect E(t)dt - \vect v_0(t)d\rho,
\end{equation}
which more explicitly underlines that, whereas the latter term is
responsible for the birth of new electrons with nonzero velocity, the
former one describes acceleration of already born electrons. The
harmonics which are typically referred to as ``Brunel harmonics''
appear due to this former term in the situation when $\rho(t)$ is
time-dependent, since if $\rho$ is constant, the corresponding
emission contains no harmonics of the driving field.
According to typical analytical theories of tunneling (for
instance \cite{yudin01a}) $\vect v_0\approx 0$, thus yielding the
Drude-like model Eq.(3). We also checked that for the low frequencies
considered in the present article and the values of $\vect v_0$,
obtained from the Wigner trajectory approach
\cite{yakaboylu14,camus17}, the term containing $\vect v_0$ is also
negligible. A figure showing relative influence of these parameters
can be found in the Supplementary. The expression
\refeq{eq:brunel-v0a} with $\vect v_0=0$ is the one shown in the
Introduction.

From the expression for $\vect J$ above, one can also
see that the average propagation direction at large times is given by
$\vect J_\infty = -\int C_0W_f(t-\tau(t)) \vect A(t) dt$. That is, the
number of electrons $dn$, born between times $t$ and $t+dt$ flying in
the direction $\phi$ determined by 
$\vect A(t)$ is given by $C_0W_f(t-\tau(t))dt$. At the same
time by the definition of $P$, the same number $dn$ is given by
$P(\phi)d\phi$. Comparing these two expressions, we obtain:
\begin{equation}
  C_0W_f(t-\tau(t))dt/d\phi = W(t)dt/d\phi = P(\phi),
  \label{eq:we}
\end{equation}
which gives us \refeq{eq:Mp}, taking into account the attoclock
mapping $t(\phi)=\mathcal M(\phi)$. In the other words: if we define a
quantity $W_e(t)$ according to \refeq{eq:Mp}, \refeq{eq:we} shows us
that $W(t)$ defined above is identical to $W_e(t)$. That is, the
effective ionization rate $W_e(t)$ obtained by mapping the angular
electron spectrum into time with $\mathcal M$ is equivalent to the
ionization rate, we observe from the Brunel harmonics, as given by the
Drude-like model \refeq{eq:brunel-v0} or \refeq{eq:drude-time}.

As we see from the derivation above for low harmonics there is a
one-to-one correspondence between wavefunction shape and the
ionization rate, reflecting the effective attoclock delay $\tau$. We
see also that the delay imaged by Brunel harmonics is the same as the
attoclock delay extracted from the electron distribution. This
attoclock delay is also in a close connection with the Wigner delay --
both are different (but closely related) measures of the group delay
of the electronic wavepacket at large times. As we have seen here, if
we consider $\tau$ as a function of time rather than a single number,
this allows to track not only delay but also asymmetry in ionization.

As a short outlook concluding this section, for higher Brunel
harmonics we can neglect neither $\vect v_0$ in \refeq{eq:brunel-v0}, nor
$T$-dependence in $\tau(t,\vect p,t-T)$, thus making harmonics
dependent on the electron dynamics even closer to the core. This
consideration is however beyond the scope of this paper.

 \subsection{Details of reconstruction of  $W(t)$ and estimation of
   errors for hydrogen and helium}

For the reconstruction of the $W_i$ in \refeq{eq:wi}, \reffig{fig:sy},
\reffig{fig:exp} we use the following procedure: for known delay
$\tau$ (which we take from \cite{eckle08}) and $W_0$ we numerically
find the values of $W_2$, $W_4$, $\delta_2$, $\delta_4$
such that the measured value of $r$ is achieved, and the center of
mass of the reconstructed distribution is located at $\tau$. This
procedure involves numerical minimization, and gives us a unique solution
for the above parameters. The value of $W_0$ was estimated from the
ionization rate at the minimum strength of our elliptical pump field.
The particular value of $W_0$ does not influences significantly the
reconstructed dynamics. In particular, despite the analytical formula
gives the symmetric ionization rate, this does not influences the
asymmetry reconstructed using our procedure. We checked this with different ionization
models (tunnel, ADK, Yudin-Ivanov) and obtained very similar results
for the rest of $W_i$.

Although the observation of the 3rd harmonic provides only limited
information, it is already sufficient to reconstruct the ionization
dynamics with a fairly good level of precision. Indeed, we observe
from our reconstruction that $|r|$ is quite small, namely $|r|=0.14$
for the Yukawa potential and $|r|=0.27$ for the Coulomb potential. It
is reasonable to expect, that the next term in the expansion of the
ionization rate is of the order of $|r|^2$ which gives us 7\%
uncertainty from the unknown contribution $W_6$ for the Coulomb
potential and 2\% for the Yukawa one. This determined error bars along
$y$-axis shown in \reffig{fig:sy}c. Besides, the uncertainty in
$\delta_6$ can lead to shifts along the time axis, which was taken
into account by adding error bars corresponding to deviations along
that axis with the amplitude equal to $|r|^2$.

In \reffig{fig:exp} the reconstruction of ionization dynamics in
helium was influenced by the impact of bound-bound transitions
(described in the text), errors to the experimental reconstruction
were estimated using analytic signal approach discussed in
Supplementary Information. In \reffig{fig:exp}, the pulse energy
measured in the experiment can not be directly converted to the peak
intensity because of the strong influence of defocusing. We used
rescaling delivered by computation of nonlinear unidirectional pulse
propagation equation for 140 $\mu$J, giving the peak intensity of 1000
TW/cm$^2$ and the waist radius 5.8 $\mu$m. Still, we note that
nonlinear propagation effects in the pump introduce certain
uncertainty. To take this into account, we introduced intensity error
bars of 10\%. The details of the numerical simulations can be found in
the Supplementary Information. This energy-to-intensity rescaling was
used for the whole intensity range.

\def\bibfont{\footnotesize}


%

\section{Acknowledgments}

I.B., A.D. and U.M are thankful to Deutsche Forschungsgemeinschaft
(DFG) (projects BA 4156/4-2, MO 850-19/2) as well as Cluster of
Excellence PhoenixD (EXC 2122, Project ID 390833453) for financial
support. O.G.K., I.A.N., N.A.P.,D.E.S. thank the support from Russian
Science Foundation, grant Nr. 21-49-00023 and National Natural Science
Foundation of China 12061131010. SS acknowledges support by the Qatar
National Research Fund (Grant NPRP 12S-0205-190047) and HPC resources
from GENCI (Grant \# A0080507594). \'{A}.J.G. acknowledges funding from
the European Union’s Horizon 2020 research and innovation programme
under the Marie Skłodowska-Curie grant agreement no. 101028938.

\section{Author contributions}

I.B. suggested the idea, developed the quantum Drude-like model,
performed simulations and analytics based on this model, performed
TDSE simulations; A.J.G performed TDSE
simulations for the two-color case; S.S.,O.G.K.,I.A.N., N.A.P., and
D.E.S. performed simulations of unidirectional propagation equations;
J.R.C.A, M.K., D.Z., L.S. and V.V., performed the experiment; I.B.,
A.J.G, A.H., F.M., L.B., S.S., U.M., A.D., T.N., M.V. and M.I.
analyzed and interpreted the results of the simulations and
experiments; all authors participated in the article formulation and
writing.

\end{footnotesize}

\end{document}